\def\3{{\ss}}
\newcommand{\beq}{\begin{equation}}
\newcommand{\eeq}{\end{equation}}
\newcommand{\beqa}{\begin{eqnarray}}
\newcommand{\eeqa}{\end{eqnarray}}

\newcommand{\krig}[1]{\stackrel{\circ}{#1}}

\documentclass[epj]{svjour}
\usepackage{latexsym}
\usepackage{graphicx}
\begin{document}
\title{The Fubini-Furlan-Rosetti sum rule and related aspects
in  light of covariant 
baryon chiral perturbation theory\thanks{Work supported in part by the DFG 
through funds provided to the  TR 16 ``Subnuclear structure of matter''.}
}
\author{
V\'eronique  Bernard \inst{1,}\thanks{
Electronic address:~bernard@lpt6.u-strasbg.fr}
\and
Bastian Kubis \inst{2,}\thanks{Electronic address:~kubis@itkp.uni-bonn.de}
\and
Ulf-G. Mei{\ss}ner \inst{2,}\inst{3,}\thanks{
Electronic address:~meissner@itkp.uni-bonn.de}
}                     
%
%
\institute{
Laboratoire de Physique Th\'eorique,
Universit\'e Louis Pasteur, F-67084 Strasbourg Cedex 2, France
\and
Helmholtz-Institut f\"ur Strahlen- und Kernphysik (Theorie),
Universit\"at Bonn, Nu{\ss}allee 14-16, D-53115 Bonn, Germany
\and
Forschungszentrum J{\" u}lich, Institut f{\" u}r Kernphysik
(Theorie),  D-52425  J{\" u}lich, Germany
}
\date{Received: date / Revised version: date}
%
\abstract{
We analyze the Fubini-Furlan-Rosetti sum rule in the framework 
of covariant baryon chiral perturbation theory to leading one--loop
accuracy and including next-to-leading order polynomial contributions.
We discuss the relation between the subtraction constants in the
invariant amplitudes and certain low-energy constants employed in earlier
chiral perturbation theory studies of threshold neutral pion photoproduction
off nucleons. In particular, we consider the corrections to the sum rule 
due to the finite pion mass and show that below the threshold they agree well with
determinations based on fixed-$t$ dispersion relations. We also discuss
the energy dependence of the electric dipole amplitude $E_{0+}$.
} 
\titlerunning{The Fubini-Furlan-Rosetti sum rule in the light of covariant CHPT}
\authorrunning{V. Bernard {\it et al.}}
\maketitle
\section{Introduction}
\label{intro}

The Fubini-Furlan-Rosetti (FFR) sum rule was derived in the sixties utilizing
the soft-pion techniques of current algebra \cite{FFR}. It relates the nucleon
anomalous magnetic moment to an integral over the invariant amplitude $A_1$
of pion photoproduction
\beq\label{FFR}
\kappa^{v,s} = \frac{8m_N^2}{e \pi g_{\pi N}} \, \int \frac{d\nu'}{\nu}
\, {\rm Im}~A_1^{(+,0)} (\nu', t=0)~,
\eeq
if one utilizes the Goldberger-Treiman relation $g_A m_N = \\F_\pi g_{\pi N}$, with
$m_N$ the nucleon mass, $g_A$ the axial-vector coupling constant, $F_\pi$ the
weak pion decay constant and $g_{\pi N}$ the strong pion--nucleon coupling 
constant. Furthermore, $\kappa^{v} = \kappa_p - \kappa_n$ and
$\kappa^s =\kappa_p + \kappa_n$ are the nucleon isovector  and isoscalar 
anomalous magnetic moment, respectively. The FFR sum
rule is exact in the chiral limit of QCD and thus all quantities appearing 
in Eq.~(\ref{FFR}) are to be understood in the limit of vanishing light quark
masses, $m_q = 0$.  The FFR sum rule has recently been reexamined in
Ref.~\cite{PDT}. In that paper\footnote{See also that paper for  references
to earlier work on the FFR sum rule.}, pion mass corrections to the sum rule were
considered (in terms of a discrepancy function $\Delta_N (\nu,t)$) and numerical 
evaluations based on a) dispersion relations and b)
input from heavy baryon chiral perturbation theory (HBCHPT) were presented. It
was pointed out that in the strict framework of HBCHPT the nucleon pole
positions are slightly moved, which leads e.g. to an incorrect curvature of
the discrepancy function for energies below the threshold. A similar behavior
due to the shift of pole or cut positions in the $1/m_N$ expansion 
was observed already in the discussion of the Compton cusp
at the opening of the pion threshold \cite{BKMS}, the spectral functions of
the nucleon isovector form factors \cite{BKMspec} or the scalar form factor
of the nucleon \cite{BL}. Note, however, that the kinematical factors
leading to the corresponding poles or cuts need
not be expanded in HBCHPT as it is discussed e.g. in
Ref.~\cite{BKMS}. Clearly, in a manifestly Lorentz-invariant formulation of
baryon CHPT such problems do not arise, see e.g. \cite{BL,KM,MZrel}. 
The purpose of this paper is two-fold: We analyze the FFR sum rule in the
framework of infrared regularization (IR) of baryon CHPT \cite{BL} 
and demonstrate that
the energy dependence of the discrepancy function is correctly given. Second,
we also take a closer look at the pion mass corrections to the sum rule,
which can only be systematically calculated in chiral perturbation theory,
and the related threshold multipoles in pion photoproduction. Our calculation
includes all terms at third order in the chiral expansion and in addition,
also the fourth order polynomial terms in the photoproduction amplitudes. Our
aim is to show that within IR baryon CHPT one can describe pion
photoproduction above and below threshold and that the dispersive
representation can indeed be used to pin down certain low--energy constants
(LECs), as suggested in \cite{PDT}. It is well-known that the chiral expansion
converges best in the unphysical region where all momenta can be very small.
Consequently, in such regions LECs can be determined to a good precision if
a corresponding dispersive representation is available. 
A more refined treatment including all
fourth order terms and fits to the existing low--energy data from MAMI will 
be relegated to a future publication.

The manuscript is organized as follows: In Sec.~\ref{sec:form} we briefly
recall the formalism of pion photoproduction and collect some results
for the pion mass corrections to the FFR sum rule derived in \cite{PDT}.
We also present an alternative way of looking at these. Sec.~\ref{sec:res}
contains the results on the FFR sum rule, the discrepancy function and the
related electric dipole amplitude $E_{0+}$ as well as the 
slopes of the P-wave multipoles
at threshold. We demonstrate that one can indeed determine LECs from the
amplitudes in the unphysical region and end with a brief outlook.

\section{Formalism}
\label{sec:form}

Consider pion photoproduction off the nucleon by real photons with $k^2=0$
\beq
\gamma (k) + N(p_1) \to \pi^a (q) + N(p_2)~,
\eeq
where $p_1\, (p_2)$ is the four--momentum of the incoming (outgoing) nucleon
($N$) and $a$ an isospin index ($a = +,0,-$). For the discussion of the FFR 
sum rule, only the isospin $0$ and $+$ channels are of relevance, i.e. the
physical channels $\gamma p \to \pi^0 p$ and $\gamma n \to \pi^0 n$.
The corresponding S-matrix is
given in terms of four invariant functions $A_i$ ($i= 1, \ldots,4$) that
depend on two kinematical variables. Throughout, we utilize the notation of
our earlier work \cite{Relphoto2} and refer to that reference for a detailed
discussion of the pertinent formalism. These invariant amplitudes can be
calculated in baryon chiral perturbation theory and have the generic form
(we do not display isospin quantum numbers and kinematical arguments)
\beq
A_i = A_i^{\rm Born} +  A_i^{\rm loop} +  A_i^{\rm ct}~,
\eeq
where the Born terms subsume the coupling to the charge and the magnetic
moment of the nucleon (note often the alternative notation of ``pole terms''
is used for these contributions -- sometimes even calculated employing the 
pseudoscalar pion-nucleon coupling). 
All further coun\-ter terms are collected in the
polynomial terms $A_i^{\rm ct}$. The non-trivial loop contributions (after
renormalization of the single nucleon properties) are collected 
in the $A_i^{\rm loop}$.  In Ref.~\cite{Relphoto2}, the $A_i$
were calculated to third order (leading loop order) in relativistic
baryon CHPT. In that formulation, a violation of the power counting through
the nucleon mass term is manifest. However, from the integral 
representations given in \cite{Relphoto2}, it is straightforward to isolate
 the so-called
infrared singular part \cite{BL} that contains the chiral long--distance
physics and leads to a one-to-one correspondence between the expansion in 
loops and small momenta/pion masses. This can e.g. be achieved by the
prescription given in \cite{BL} which we also will employ. Symbolically,
it reads $\int_0^1 dx \to (\int_0^\infty - \int_1^\infty) dx = I + R$,
with $I$ and $R$  the infrared singular (irregular) and the regular part, 
respectively.  For the study of the FFR sum rule
and related aspects, we are interested in the amplitudes at small energies
and momentum transfer and thus use the variables $\nu = (s-u)/4m_N$ and
$\nu_B = - (s+u-2m_N^2)/4m_N$, which are odd and even under crossing $s
\leftrightarrow u$ (for further notation, see \cite{Relphoto2}). From the
crossing properties of the $A_i$ and their low-energy properties as detailed
in \cite{Relphoto2}, one derives the following representation for the
polynomial pieces (note in particular the low-energy theorem for $A_1$ 
\cite{Relphoto1} that forbids a constant term) 
\beqa\label{Aexpand}
A_1^{\rm ct} &=& a_1^1 \nu^2 + a^2_1 \nu_B + \ldots ~,\nonumber\\
A_2^{\rm ct}  &=& a_2^0 + \ldots ~,\nonumber\\
A_3^{\rm ct}  &=& a_3^1 \nu +  \ldots ~,\nonumber\\
A_4^{\rm ct}  &=& a_4^0 + \ldots ~.
\eeqa 
At third order in the chiral expansion, only the leading term of $A_4$
contributes \cite{Relphoto2}, 
whereas all other terms written down in Eq.~(\ref{Aexpand})
start at ${\cal O}(q^4)$. The mapping between these subtraction constants
and the low-energy constants (LECs) used in the heavy baryon calculations
\cite{HBphoto1,HBphoto2,HBphoto3} is given in the appendix. We remind the
reader that although the contribution of $A_1^{\rm ct}$ formally starts
as $M_\pi^2$ at threshold, in the chiral limit the expansion coefficients
are singular leading to the famous contribution to the low-energy theorem
for the threshold value of $E_{0+}$ at next-to-leading order in the pion
mass expansion \cite{Relphoto1}. We note that  $A_2^{\rm ct}$ only feeds 
into the P-wave slope  $\bar P_2$ and into the D-wave at threshold
in such a way that
it cancels in the FFR sum rule at finite pion mass. Therefore, the subtraction constant
$a_2^0$ has to  be determined completely independently of the FFR sum rule, say
by fitting to the slope of $P_2$ at threshold. In the following,
we use the third order IR representation of the pertinent one-loop graphs
(which contains an infinite series of $1/m_N$ corrections in the heavy baryon
framework) but also use the fourth-order subtraction constants displayed
in Eq.~(\ref{Aexpand}). Based on that representation, we attempt a
simultaneous description of the $\nu$-dependence of the FFR discrepancy 
function $\Delta_p (\nu, t_{\rm thr})$ (as defined below), 
the energy dependence of the electric 
dipole amplitude $E_{0+}$ for neutral pion production off protons and the 
P--wave threshold slopes as extracted in \cite{E0pdata}.  Similarly precise
information is not available for the neutron, therefore in the following we
mostly concentrate on the proton.

In Ref.~\cite{PDT}, the FFR sum rule was considered for finite mass pions 
and the following representation in terms of a discrepancy function $\Delta_N$
was derived:
\beqa\label{disp}
&&\krig\kappa_N \tau_3 + \Delta_N (\nu, t_{\rm thr})\nonumber\\ 
&& \quad = \frac{4m_N^2}{e \pi g_{\pi N}} \, 
\int_{\nu_{\rm thr}}^\infty \frac{d\nu'}{\nu}
\,\frac{\nu' \, {\rm Im}~A_1^{(N,\pi^0)} (\nu', t=t_{\rm thr})}{{v'}^2 -
  \nu^2}~,\nonumber \\
&& \Delta_N (\nu, t=t_{\rm thr}) \nonumber \\
&& \quad =\frac{2m_N^2}{e  g_{\pi N}} \,\left( A_1^{\rm loop}(\nu, t=t_{\rm
    thr}) +  A_1^{\rm ct}(\nu, t=t_{\rm  thr}) \right).
\eeqa
A few comments on this equation are in order. First, the
left-hand-side of the FFR sum rule gives the anomalous magnetic moment in the
chiral limit, $\kappa_N = \krig\kappa_N + {\cal O}(m_q^{1/2})$. Therefore, if
one uses the physical value of $\kappa_N$ as in Ref.~\cite{PDT}, 
one must include the corresponding
loop and counter term corrections into the discrepancy function. However,
for studying the pion mass corrections to the FFR sum rule, it is more
appropriate to  work with the chiral limit values of $\kappa_p$ and $\kappa_n$,
as discussed below.
Second, in the chiral limit, $t \to 0$. This value can, of course, not
be achieved in the physical world. We follow \cite{PDT} and present our
results at the minimal (threshold) value of $t$, $t_{\rm thr} = -M_\pi^2 /
(1 + M_\pi/m) = -0.016\,$GeV$^{-2}$.
Third, the Goldberger-Treiman relation is no longer exact at finite pion
mass, to the order we are working, it takes the form \cite{GSS,FMS}
\beq
\label{gpin}
\frac{g_{\pi N}}{m_N} = \frac{g_A}{F_\pi}
\left( 1 - \frac{ 2 M_\pi^2}{g_A} \bar d_{18} \right)\,,
\eeq
with $\bar d_{18}$ a LEC.
As long as one only works at the physical pion mass, this effect is taken
care of by utilizing the physical value for the pion--nucleon  coupling and
the nucleon mass. If one, however, also wants to study the pion mass
dependence of $\Delta_N$, as will be done here, one explicitly has to
include this pion mass dependence. However, this effect only shows up in
the terms cubic in the pion mass, which will not be considered in detail here
(for a more detailed discussion of this topic, see e.g. \cite{EGM}).
Note this pion mass dependence can only be systematically calculated in
chiral perturbation theory and, eventually, in lattice QCD. 

We have explicitly worked out  the quark mass expansion of the discrepancy
function at threshold.  For the proton, it takes the form (modulo chiral logs)
\beq\label{DeltapMpi}
\Delta_p (\nu=\nu_{\rm thr}, t = t_{\rm thr}) 
= \alpha_p M_\pi + \beta_p M_\pi^2
+ \ldots
\eeq
with 
\beqa
\alpha_p & = & \frac{ m_N}{16  F_\pi^2}\,\left(1 - 
\frac{4 g_A^2}{3 \pi} \right)~,\nonumber \\
\beta_p &=& \frac{1}{8 \pi^2 F_\pi^2}
\biggl[ -3 - \frac{\pi^2}{8} 
 + \bigl( 7 -\frac{1}{2}\krig\kappa_p +\frac{1}{2}\krig\kappa_n
 \bigl) \ln\frac{M_\pi}{m_N}\biggr]\nonumber \\
&-& \frac{g_A^2}{48 \pi^2 F_\pi^2}\biggl[-\frac{85}{3} - \pi 
+ \frac{5}{3} \krig\kappa_p  - \frac{11}{3}\krig\kappa_n \nonumber\\
&& \qquad\qquad\qquad\qquad 
+ (5+ 7\krig\kappa_p - \krig\kappa_n )\ln\frac{M_\pi}{m_N}\biggr]\nonumber \\ 
 &+&  \frac{\tilde{c}_4}{6\pi^2F_\pi^2} 
\left(-\frac{10}{3}  + \frac{3\pi^2}{8} + \frac{5}{2} \ln\frac{M_\pi}{m_N}
\right)
\nonumber\\
 &-& 8m_N (2 e_{105}+e_{106}) 
+\frac{2m_N F_\pi}{g_A} 
\left( a_1^1 - \frac{a_1^2}{2m_N} \right)   ~,
\eeqa
where $\tilde{c}_4 = m_N \, c_4$ and $c_4 = 3.4\,$GeV$^{-1}$ \cite{BM}.
The LECs $e_{105}$ and $e_{106}$ from the dimension four chiral pion-nucleon
Lagrangian contribute to the proton and neutron anomalous magnetic moment
at next-to-leading loop order (for a detailed discussion, see \cite{KM}). 
Their values are discussed below.
Further, we have set the scale of dimensional regularization equal to the
nucleon mass, $\lambda = m_N$.
Of course, $\Delta_p$ vanishes in the chiral limit. Notice the absence
of chiral logs in the terms linear in the pion mass.
The representation of $\Delta_p$ given Eq.~(\ref{DeltapMpi}) is exact to fourth order
in the chiral expansion since it can be reconstructed from the HBCHPT results
obtained in \cite{HBphoto1,HBphoto2,HBphoto3}. In addition,
to arrive at these results, we had to include the small contribution from the
slope of the D-wave combination $D = M_{2+}-E_{2+}-P_{2-}-E_{2-}$, that is
$\bar D = D/q^2$ at threshold. It is obtained from the invariant functions by
\beq
D(s) = \frac{5}{16} \, \int dx ~(x^4 - 2x^2 + 1) \, {\cal F}_4(s,x)~,
\eeq
with ${\cal F}_4(s,x)$ a combination of $A_{2,3,4}$, see \cite{Relphoto2}.
This contribution has not been calculated before.
Note further that Eq.~(\ref{DeltapMpi})
includes the terms  that renormalize $\krig\kappa_p$ to its physical
value, $\kappa_p$, since we identify the left-hand-side of the FFR with the
anomalous magnetic moment in the chiral limit.
The pion mass expansion
of $\Delta_n (\nu=\nu_{\rm thr}, t = t_{\rm thr})$ looks very similar, we
find
\beqa
\alpha_n & = & \frac{ m_N}{16  F_\pi^2}\,\left(1 - 
\frac{4 g_A^2}{3 \pi}  \right)~,\nonumber \\
\beta_n &=& -\frac{1}{8 \pi^2 F_\pi^2}
\biggl[ \frac{7}{9} + \frac{\pi^2}{8} 
+ \bigl(-\frac{13}{3}  + \frac{1}{2} \krig\kappa_p 
- \frac{1}{2} \krig\kappa_n \bigr) \ln\frac{M_\pi}{m_N}\biggr]
\nonumber \\
&-& \frac{g_A^2}{48 \pi^2 F_\pi^2}\biggl[-\frac{76}{3} - \pi 
+ \frac{11}{3} \krig\kappa_p  - \frac{5}{3}\krig\kappa_n \nonumber\\
&& \qquad\qquad\qquad\qquad 
+ (-1 + \krig\kappa_p  -7\krig\kappa_n )\ln\frac{M_\pi}{m_N}\biggr]\nonumber
 \\ 
 &-&  \frac{\tilde{c}_4}{6\pi^2F_\pi^2} 
\left(-\frac{10}{3}  + \frac{3\pi^2}{8} -\frac{5}{2} \ln\frac{M_\pi}{m_N}\right)
\nonumber\\
 &+& 8m_N (2 e_{105}-e_{106})  
+\frac{2m_N F_\pi}{g_A} 
\left( a_1^{1,n} - \frac{a_1^{2,n}}{2m_N} \right)   ~,
\eeqa
where  the subtraction constants refer to the neutron amplitude
$\gamma n \to \pi^0 n$ as denoted by the superscript $n$.
\section{Results}
\label{sec:res}

First, we must fix our input parameters. We use $F_\pi = 92.4\,$MeV, 
$M_{\pi^+} = 139.57\,$MeV, $M_{\pi^0} = 134.97\,$MeV, 
$m_p = 938.27\,$MeV, $m_n = 939.57\,$MeV,
$g_{\pi N} =13.4$, $\kappa_p = 1.793$, $\kappa_n = -1.913$.
To the order we are working, the anomalous magnetic moment of the
proton and the neutron are given in terms of two LECs from the
dimension two chiral Lagrangian commonly denoted $c_6$ and $c_7$ and
loop corrections that start with terms of order $M_\pi^3$ \cite{BKMlec}.
From that one can read off the chiral limit values for the proton and
the neutron anomalous magnetic moments,
\beq\label{kappakringel}
\krig\kappa_p = 2.37~, ~~ \krig\kappa_n = -2.84~,
\eeq
in nuclear magnetons.  For comparison, at third order in HBCHPT, we
have $\krig\kappa_p = 2.85$,  $\krig\kappa_n = -2.98$ \cite{BKMlec}.

The overall best fit to simultaneously describe  the proton discrepancy
function, the electric dipole amplitude in the threshold region and the
three P-wave slopes is obtained with the following polynomial contribution
to the invariant functions 
\beqa\label{bestfit} 
A_1^{\rm ct}  (\nu,\nu_B) &=& (191.3 \nu_B + 220 \nu^2)~\mbox{GeV}^{-2} ~, 
\nonumber \\
A_2^{\rm ct}  (\nu,\nu_B) &=& -54.9~\mbox{GeV}^{-4} ~,  \nonumber \\
A_3^{\rm ct}  (\nu,\nu_B) &=& -155 \nu~\mbox{GeV}^{-3} ~, \nonumber \\
A_4^{\rm ct}  (\nu,\nu_B) &=& 181.5~\mbox{GeV}^{-3}~.
\eeqa
While these coefficients appear large at first glance, in the appendix we
show that  they match quite nicely the corresponding LECs determined in fits
to neutral pion photoproduction differential cross sections and the photon 
asymmetry. Since the LECs can be understood to  a good precision in terms of
resonance saturation (excitation of the $\Delta (1232)$ and of vector mesons, see
e.g. \cite{HBphoto1}), the numbers appearing in Eq.~(\ref{bestfit}) are indeed
of natural size. The resulting
threshold values for $E_{0+}$ and the $\bar P_i$ ($i=1,2,3$) are 
\beqa\label{thrval}
E_{0+} &=& -1.19 \,~~[-1.23 \pm 0.08 \pm 0.03]~,\nonumber \\
\bar P_1 &=& \phantom{-}9.67 \,~~ [9.46  \pm 0.05 \pm 0.28]~,\nonumber\\
\bar P_2 &=& -9.6 \,~~ [-9.5  \pm 0.09 \pm 0.28]~,\nonumber\\
\bar P_3 &=&  \phantom{-}11.45 \,~[11.32  \pm 0.11 \pm 0.34]~,
\eeqa 
in the conventional units of $10^{-3}/M_{\pi^+}$ and $10^{-3}/M_{\pi^+}^2$, 
respectively. The experimental numbers in the square brackets are from \cite{E0pdata}.
We note that  $\bar P_2$ is obtained by adjusting the subtraction
constant $a_2^0$. The corresponding D-wave slope is 
\beq
\bar D = 0.66 \cdot 10^{-3}/M_{\pi^+}^3~,
\eeq
to be compared with 0.96 (0.92) from the MAID03 analysis (the dispersive
analysis of Ref.~\cite{HDT}).

Let us look at these results in more detail.  For the pion mass
correction to the proton FFR sum rule, we find 
\beq\label{FFRres}
\krig\kappa_p + \Delta_p (\nu = \nu_{\rm thr},t = t_{\rm thr}) = 2.18~
\eeq
which should be compared with the left-hand-side of the sum rule, i.e. 
the proton anomalous magnetic moment in the
chiral limit, cf. Eq.~(\ref{kappakringel}).
Thus, within this framework, the FFR sum rule is
fulfilled within 8\%, which is of the expected size  since pion mass effects
are proportional to the small parameter $\mu = M_\pi/m_N \simeq 1/7$. Note again
that this way of looking at the FFR differs from what was done in \cite{PDT},
where the left-hand-side of the FFR was identified with the physical value of
the proton anomalous magnetic moment and the discrepancy function defined
there thus differs from ours by terms $\sim \krig\kappa_p - \kappa_p$.
 
\begin{figure}[t]
\vspace{-6mm}
\centerline{\includegraphics*[width=6cm,angle=270]{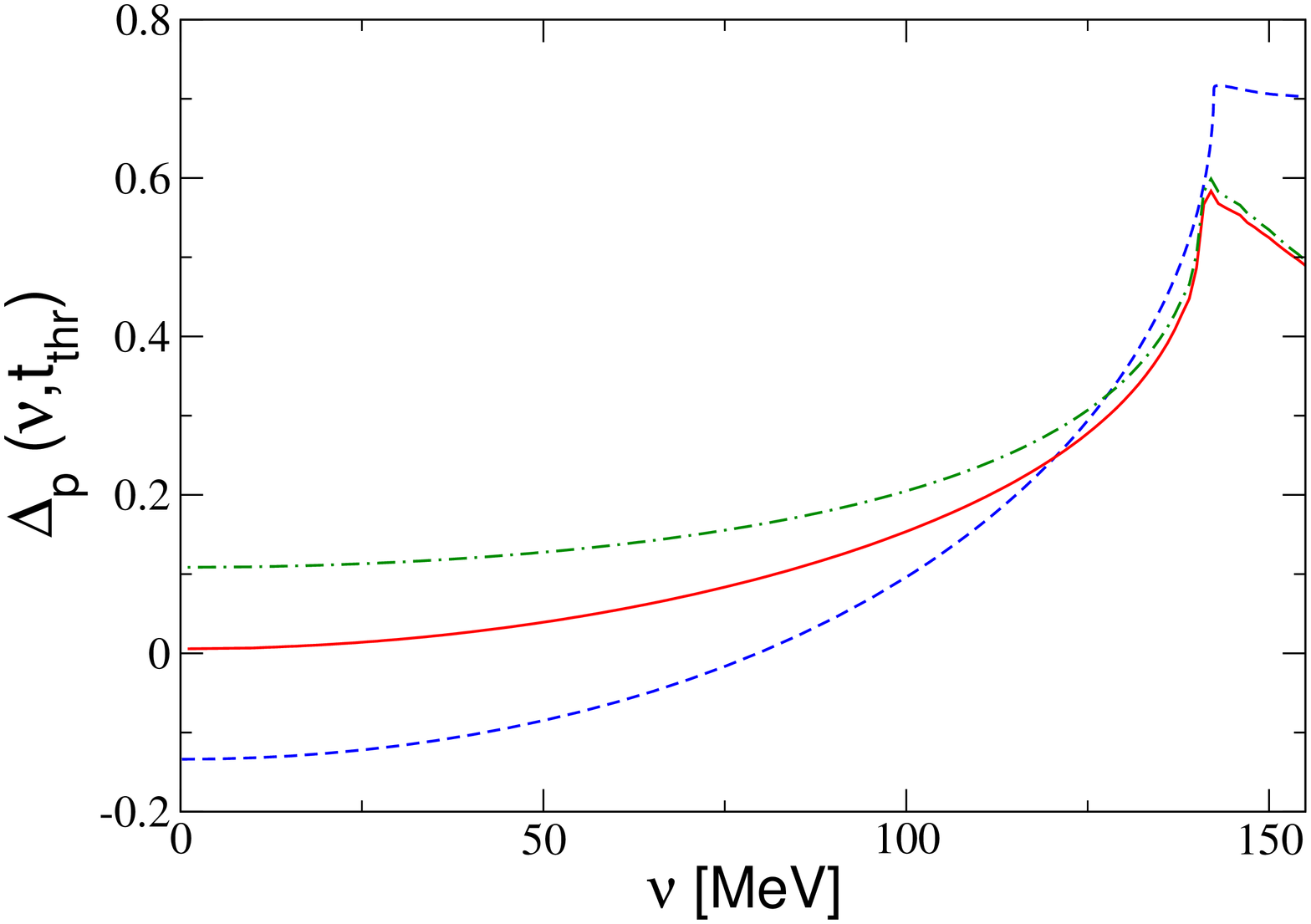}}
\vspace{-4mm}
\centerline{\includegraphics*[width=6cm,angle=270]{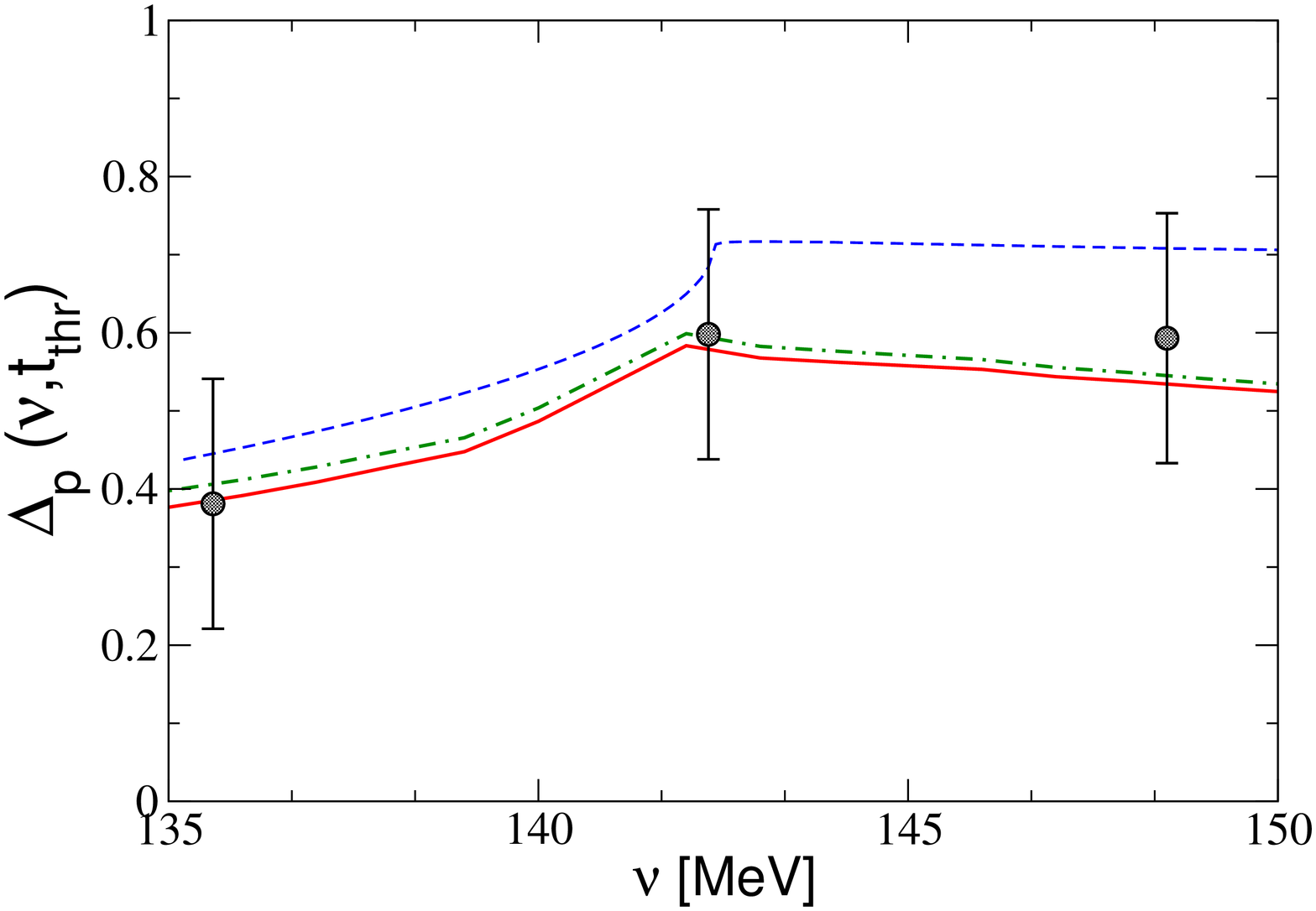}}
\caption[fig2]{\label{fig2}
The discrepancy function $\Delta_p$ for the proton. Solid line:
IR baryon CHPT. Dot-dashed line: third order relativistic CHPT
\protect\cite{Relphoto2}. Dashed line: Dispersive result
based on the MAID model, from \protect\cite{PDT}.
Upper panel: $\nu$ in the range from 0 to 155~MeV.
lower panel: Threshold region (135~MeV$\leq \nu \leq$150~MeV).
The data points are calculated with the S- and P-wave multipoles
from Ref.~\protect\cite{E0pdata} (based on the formalism developed in
\protect\cite{PDT}).
}
\end{figure}

Next, we display the resulting discrepancy function for the proton as a function
of $\nu$ at fixed $t = t_{\rm thr}$ in Fig.~\ref{fig2} (for better comparison
with Ref.~\cite{PDT}, this discrepancy function is taken with respect to the
physical value of the proton anomalous magnetic moment).
At $\nu = 0$, the discrepancy functions has indeed an extremum (as
pointed out in \cite{PDT}) and it increases with increasing $\nu$.
We also find the pronounced cusp effect at $\nu = \nu_{\rm thr}$, as
it is expected.  In contrast to the prediction based on the MAID model
(dashed curved), we do not observe any zero crossing for the values
of the subtraction constants given in Eq.~(\ref{bestfit}). Naturally, within
our approach we should have a band rather than a line but we relegate a
detailed error analysis to a later work, when we will also fit to all 
data of  neutral pion photoproduction in the threshold region. Interestingly,
the third order relativistic result from \cite{Relphoto2} (dot-dashed line) is
very close to the IR result including the fourth order polynomial pieces.
We also note that the incorrect behavior at $\nu \simeq 0$ observed in the
HBCHPT calculation is of course not present in our manifestly covariant
calculation, as it was suggested already in \cite{PDT}.
Note furthermore that we do not show $\Delta_P (\nu, t_{\rm thr})$ for values 
of $\nu > 155\,$MeV because at $\nu \simeq 170\,$MeV, a steep rise due to the
$\Delta(1232)$ resonance sets in \cite{PDT}.  In the lower panel of 
 Fig.~\ref{fig2} we focus on the threshold (cusp) region. We see that
the relativistic and IR predictions are in good agreement with the MAID
model and also with the data reconstructed form the multipoles of
Ref.~\cite{E0pdata} (for details, see again \cite{PDT}). 

We briefly discuss the pion mass dependence of $\Delta_{p,n}$; for this
purpose, we switch back to the full fourth order representation
obtained from earlier heavy baryon results. Consider
the proton. We find for the coefficients in Eq.~(\ref{DeltapMpi})
\beqa
\alpha_p &=& 2.19~{\rm GeV}^{-1}~, \nonumber\\
\beta_p  &=& (-5.0 -17.3)~{\rm GeV}^{-2} = -22.3~{\rm GeV}^{-2}~,
\eeqa
where we have used the physical values for $g_A$ and $F_\pi$
and the second term in $\beta_p$ is the contribution from the
dimension four operators with the LECs 
$e_{105}$ and $e_{106}$. These counterterms determine the slope of 
$\kappa_v$ and $\kappa_s$ in the soft pion limit. 
Their values have been determined by using the
fourth order formula for the anomalous magnetic moment from \cite{KM}
as chiral extrapolation functions for the lattice QCD data of \cite{FFlat}.
We get $e_{105} \simeq 0.45 \, ,\, e_{106} \simeq 1.4$ at $\lambda = m_N$, which 
are of natural size. Since these are only very rough fits to 
the trend of the lattice data at too high pion masses, 
we refrain from assigning a theoretical uncertainties to these numbers. 
Note that in case of the coefficient $\beta_p$, we have large
cancellations between the loop and the counterterm contributions  proportional
to the LECS $c_i$, which are $-21.2\,$GeV$^{-2}$ and $16.2\,$GeV$^{-2}$, 
respectively, for $\lambda = m_N$. We also stress that the
first two terms in the quark mass expansion of $\Delta_p$ at threshold
give $-0.13$, which is to be compared with the full calculation that
gives $-0.19$, cf. Eq.~(\ref{FFRres}) (where the contribution from the term
$\sim 2e_{105} + e_{106}$ is $-0.34$).
For the neutron, we have no determination of the photoproduction
counter terms and thus 
we can only give $\alpha_n  = \alpha_p =2.19\,$GeV$^{-1}$  
and $\beta_n^{\rm loop} = -18.23 \,$GeV$^{-2}$.
This value is comparable to the one of the proton. The contribution from
the operators $\sim e_{105}, e_{106}$ is somewhat smaller than for the proton
since for the neutron they appear with a different relative sign.

\begin{figure}[tb]
\vspace{-6mm}
\centerline{\includegraphics*[width=7cm,angle=270]{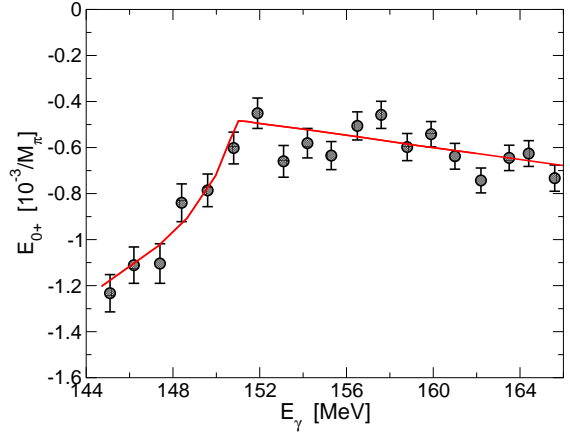}}
\caption[fig1]{\label{fig1}
The real part of the electric dipole amplitude $E_{0+}$ in the threshold
region. The solid line is the theoretical prediction as discussed in the
text and the data are from Ref.~\protect\cite{E0pdata}. 
}
\end{figure}

Also well described is the energy dependence of the electric dipole
amplitude $E_{0+} (E_\gamma)$ shown in Fig.~\ref{fig1} (with $E_\gamma
= \nu-(t -M_\pi^2)/4m_N$). This description of the data is as good as the
complete fourth-order HBCHPT calculation (see e.g. Fig.~3 in \cite{E0pdata}).
This is not  surprising since the third-order IR calculation generates
all fourth-order HB corrections with fixed coefficients (from the expansion
of the Dirac propagator) and these are the only fourth order loop 
contributions since loops with one insertion proportional to 
the dimension two LECs $c_i$ do not contribute. Related to this is the
observation that  most of the non-trivial energy-dependence is given by the 
generalized cusp function \cite{HBphoto1,HBphoto3} 
\beq
E_{0+} (E_\gamma)= a + b \sqrt{1 - \left(\frac{E_\gamma}
{E_{\gamma}^{\rm thr}}\right)^2}~,
\eeq
where the parameter $b$ is proportional to the charge exchange scattering
length $a(\pi^0 p \to \pi^+ n)$ (see also the detailed discussion in
\cite{aron1,aron2}).

As noted before, similarly precise information on the neutron is not available.
The predictions for the threshold multipoles are based on resonance saturation,
the only experimental information is from the SAL experiment on 
$\gamma d \to \pi^0 d$ \cite{SAL} that is consistent with the CHPT prediction 
$|E_{0+, \rm thr}^{\pi^0n}| > |E_{0+, \rm thr}^{\pi^0p}|$ \cite{BBLMvK}.
We can fit to the energy-dependence of $A_1^n$ as predicted by the MAID model
and the threshold multipoles as predicted by CHPT, but in the absence of more information
on the energy dependence of e.g. the neutron electric dipole amplitude and any
experimental verification of the predictions for the P--wave slopes based
on resonance saturation, the resulting numbers are highly model-dependent and 
we thus refrain from showing them here.

\section{Summary and outlook}
\label{sec:out}

In this paper, we have studied the FFR sum rule in the framework of
covariant baryon chiral perturbation theory, extending some of the
work presented in \cite{PDT}. We have worked out the
loop corrections to third order in the chiral expansion, corresponding
to the leading loop contribution, supplemented by the polynomial pieces
up-to-and-including fourth order. Since the (sub-)threshold energies
considered here are small, such a procedure is justified and would
only lead to a readjustment of the subtraction constants defined
in Eq.~(\ref{Aexpand}) if one includes also the fourth order
loop graphs. We have shown that within
this framework one can achieve a good description of the energy dependence
of the discrepancy function for the proton, defined in Eq.~(\ref{disp}),
together with the energy dependence of the electric dipole amplitude in the
threshold region, cf. Fig.~\ref{fig1}, and the P-wave slopes at threshold,
see Eq.~(\ref{thrval}). The corresponding subtraction constants collected
in Eq.~(\ref{bestfit}) can be matched to the low-energy constants determined
previously in HBCHPT studies of threshold neutral pion photoproduction and their
resulting values are of natural size (as detailed in the appendix). 
We find that the finite pion mass corrections to the FFR 
sum rule for the proton are  small, of the order of 
8~\% (cf. Eq.~(\ref{FFRres})). As shown in Fig.~\ref{fig2}, the $\nu$-dependence
of the discrepancy function for the proton has the proper behavior and agrees
with the result of the MAID model. The unphysical behavior observed at $\nu \simeq 0$
in the heavy baryon scheme \cite{PDT} is absent in a covariant formulation as presented
here. It is also interesting to note that the third order relativistic calculation
free of counter terms from \cite{Relphoto2} also gives a good description of 
$\Delta_p (\nu, t_{\rm thr})$. 
These findings corroborate the conjecture made in Ref.~\cite{PDT} that one
can use the dispersive representation of the invariant amplitudes in the unphysical
region to pin down low-energy constants of chiral perturbation theory (see 
also the appendix). It will be interesting to perform a 
complete fourth-order calculation and fit the corresponding LECs to the existing
unpolarized and polarized
threshold data of the reaction $\gamma p \to \pi^0 p$. This should further sharpen
the conclusions made here. Work along such lines is under way \cite{BKMfull}.

\section*{Acknowledgments}
We thank Dieter Drechsel for interesting us  in this problem.
We are grateful to Lothar Tiator for supplying us with the
results of Ref.~\cite{PDT}.
This research is part of the EU Integrated Infrastructure 
Initiative Hadron Physics Project
under contract number RII3-CT-2004-506078.

\appendix
\section{Subtraction and low-energy constants}

Here, we give the mapping between the commonly employed counter terms
of the heavy baryon approach and the subtraction constants defined
in Eq.(\ref{Aexpand}). 
To third order, one has only one P--wave counter term (with the LEC $b_p$)
that feeds into the multipole $P_3$ \cite{HBphoto1},
\beq
a_4^0 = 4 \pi \, b_P~.
\eeq
At fourth order, there are two S--wave counter terms (which in 
the threshold region essentially act as one constant) \cite{HBphoto1}. 
The corresponding LECs $a_1$ and $a_2$  
are given by the following combinations of subtraction constants:
\beqa\label{a12}
12 \pi\, a_1 &=& -\left(a_2^0 + \frac{a_4^0}{m_N} \right) ~, \nonumber \\
12 \pi\, a_2 &=& \left(3a_3^1 + a_2^0 + 3a_1^1 - \frac{3}{2}\frac{a_1^2}{m_N} 
+ \frac{5}{2}\frac{a_4^0}{m_N}\right)~.
\eeqa
Note that in the sum $a_1 + a_2$, the contribution from $A_2$ cancels (as
noted earlier) and that these relations are to be taken at $\lambda = m_N$.
Similarly, at fourth order there are two independent counter terms 
(with the LECs $\xi_1$ and $\xi_2$)  that
modify the P-waves $P_1$ and $P_2$ \cite{HBphoto3}
\beqa\label{xi12}
\frac{a_1^2}{2} -m_N a_3^1 - a_4^0 &=& \frac{g_A}{16\pi^2F_\pi^3}~ \xi_1~, \nonumber\\
\frac{a_4^0}{2} +  m_N a_3^1 - m_N a_2^0
&=&  \frac{g_A}{16\pi^2F_\pi^3} ~\xi_2~.
\eeqa
At first glance, one might conclude from Eqs.(\ref{a12},\ref{xi12}) that there
is a mismatch in the number of subtraction constants and counter terms.
Note, however, that various  subtraction constants feed into the
S- and the P-waves so that finally only two independent structures remain for
the S-wave and two for the P-waves.

It is interesting to compare the numbers derived from Eq.~(\ref{bestfit})
with the earlier determinations of these counter terms in the HBCHPT
framework. Note, however, that we did not include all fourth order
loop corrections here, so that the values of the subtraction constants
effectively subsume some of these effects. This is not the case for
the LEC $b_P$ since it already appears at third order. Our value for
$a_0^4$ translates into $b_P = 14.4\,$GeV$^{-3}$. This compares well
with the third order fits of Ref.~\cite{HBphoto1}, 
$b_P = (15.8 \pm 0.2)\,$GeV$^{-3}$, and of  Ref.~\cite{HBphoto2},
$b_P = 13.0\,$GeV$^{-3}$. Note that the corresponding value in
\cite{HBphoto3} comes out smaller due to additional loop effects.
Consider next the LECs contributing to the electric dipole amplitude in
the threshold region. We get $4\pi m_N (a_1+a_2)
= 56.2\,$GeV$^{-3}$  from the constants in
Eq.~(\ref{bestfit}) compared to 31.8~GeV$^{-3}$,  77.8~GeV$^{-3}$ 
and  71.2~GeV$^{-3}$ from \cite{HBphoto1}, \cite{HBphoto2} and 
\cite{HBphoto3}, respectively.  Individually, we have $a_1 =
-3.67\,$GeV$^{-3}$ and $a_2 = 8.43\,$GeV$^{-3}$, which is different from
but comparable in size to the
free and resonance fits to the various sets of Mainz and Saskatoon
data (compare e.g. table~1 in \cite{HBphoto3}). As it was already stressed
in these earlier papers, the LECs $a_1$ and $a_2$ can not be well determined
individually from fits to the data in the threshold region. 
Next, consider the P-wave $P_1$.
The determination of $\xi_1 = 16.6$ in \cite{HBphoto3} translates into
$a_1^2/2 - m_N a_3^1 - a_4^0 = 175.7\,$GeV$^{-3}$, 
which is sizably larger than the value
of 59.8 obtained from  Eq.~(\ref{bestfit}). This is expected since in 
 \cite{HBphoto3} it was shown that there are large cancellations between
fourth-order loop and counter term contributions, which we represent by the
polynomial term only. This discrepancy is even more pronounced for
the  combination of subtraction constants that
can be obtained from $\bar P_2$. Utilizing $\xi_1 = -19.7$ from \cite{HBphoto3}
and the values from Eq.~(\ref{bestfit}), we obtain
$a_4^0/2  + m_N a_3^1 - m_N a_2^0 = -208.5$~GeV$^{-3}$ and $-3.3$~GeV$^{-3}$, 
respectively. This shows that the cancellations between the fourth order loop
and counter term contributions are even stronger in $P_2$ than in $P_1$.


\end{document}